\documentstyle[preprint,aps,pra,epsf]{revtex}

\newcommand{\be}{\begin{equation}}
\newcommand{\ee}{\end{equation}}
\newcommand{\bea}{\begin{array}}
\newcommand{\ea}{\end{array}}
\newcommand{\beqa}{\begin{eqnarray}}
\newcommand{\eeqa}{\end{eqnarray}}
\newcommand{\bean}{\begin{eqnarray*}}
\newcommand{\eean}{\end{eqnarray*}}
\newcommand{\eqn}[1]{(\ref{#1})}

\newcommand{\del}{\partial}
\newcommand{\nn}{\nonumber}

\def\up#1{\leavevmode \raise.16ex\hbox{#1}}

\newcommand{\gapproxeq}{\lower .7ex\hbox{$\;\stackrel{\textstyle >}{\sim}\;$}}
\newcommand{\lapproxeq}{\lower .7ex\hbox{$\;\stackrel{\textstyle <}{\sim}\;$}}

\newcounter{appendice}

\def\thebibliography#1{{\bf REFERENCES\markboth
 {REFERENCES}{REFERENCES}}\list
 {[\arabic{enumi}]}{\settowidth\labelwidth{[#1]}\leftmargin\labelwidth
 \advance\leftmargin\labelsep
 \usecounter{enumi}}
 \def\newblock{\hskip .11em plus .33em minus -.07em}
 \sloppy
 \sfcode`\.=1000\relax}

\begin{document}
\title{
\hfill {
\small{$\stackrel
{\rm\textstyle DSF-57/97}
{
\rm\textstyle quant-ph/9802030
}
$}}
\\[1truecm] 
Time--Dependent  Invariants and Green Functions in the Probability 
Representation of Quantum Mechanics} 
\author{V. I. Man'ko\thanks{On leave from Lebedev Physical 
Institute, Moscow, Russia}, L. Rosa and  P. Vitale }
\address{ Dipartimento di Scienze Fisiche, Universit\`a di Napoli,\\
Mostra d'Oltremare, Pad.19, I-80125, Napoli, Italy; \\
INFN, Sezione di Napoli, Napoli, ITALY.\\
\small e-mail: \tt manko,rosa,vitale@na.infn.it } 
\maketitle
\begin{abstract}
In the  {\it probability representation} 
of quantum mechanics, quantum states are represented by 
a classical probability distribution, the marginal 
distribution function (MDF), whose time dependence is governed by a 
{\it classical} evolution equation. We find and explicitly solve, for 
a wide class of Hamiltonians, new equations for the Green's function of 
such an equation, the so--called classical propagator. We elucidate the 
connection of the classical propagator to the quantum propagator for 
the density matrix and to the Green's function of the Schr\"odinger 
equation. Within the new description of quantum mechanics we give a 
definition of coherence solely in terms  of properties of the MDF and 
we test   the new definition recovering well known results. As an 
application, the forced parametric oscillator is considered . Its 
classical and quantum propagator are found, together with the MDF for 
coherent and Fock states.
\end{abstract}

\bigskip


\newpage
\section*{Introduction}
Quantum states are usually described in terms of wave functions
\cite{schr} (for pure states) or by means of the density matrix
\cite{landau}, \cite{vonneuman} (for mixed states). Nonetheless, since
the beginning of quantum mechanics there have been attempts of
understanding  the notion of quantum states in terms of a classical
approach \cite{debroglie,bohm}. 
Due to the Heisenberg 
principle, it is not possible to introduce a joint distribution function for
both coordinates and momenta, while these joint probability 
distributions are the main tool to describe physical states in 
classical statistical mechanics. It is such kind of problems which has
brought to the introduction of the so called quasi--probability
distribution functions, such as the Wigner function \cite{wig32}, the
Husimi function \cite{husimi} and the Glauber--Sudarshan function
\cite{glauber63}, \cite{sud63}, later on unified into a one--parametric
family \cite{cahill}.  Despite their wide use in quantum theory and
their fundamental r\^ole in clarifying the link among classical and
quantum aspects, these quasi--probability distributions cannot play the
r\^ole of classical distributions since, either they allow for negative
values (like the Wigner function) or they do not describe 
distributions of measurable variables. A formulation of quantum 
mechanics which is very similar to the classical stochastic approach 
has been presented by Moyal \cite{moyal}. But, the evolution equation 
suggested by Moyal was an equation for a quasi--probability 
distribution function (the Wigner function) and not for the 
probability. 

In \cite{tom3,tom33} it was recently suggested to consider
quantum dynamics as a classical stochastic process described namely by
a probability distribution:   the so called  {\it marginal distribution
function} (which was discussed in a general context in \cite{cahill}), 
associated to  the position coordinate, $X$,  taking values 
in an ensemble of reference frames in the phase space. Such a classical
probability distribution is shown to completely describe quantum states
\cite{tom1,tom2}. The approach of \cite{tom3}-\cite{tom2} 
was developed both for quadrature observables \cite{manko1}-\cite{new} 
and for spin \cite{olga}. 

Within this approach the notion of ``measuring a
quantum state" provides the usual ``optical tomography approach"
\cite{berber,vogrisk,raymer} and its extension called the ``symplectic
tomography" formalism \cite{tom1,tom2}. Both allow for  an explicit
link between the MDF and the Wigner function
or the density matrix, in other representations. In this way, starting
from the evolution equation for the density matrix, an evolution
equation of the Fokker-Plank type for the marginal distribution
function is obtained \cite{tom3,tom33}. Such an equation allows for an 
independent
definition of the marginal distribution function. Thus it may be
assumed as the starting point for an alternative but equivalent
formulation of quantum mechanics in what we call the {\it probability
representation}, or classical-like description of quantum mechanics.
In such a scheme, it plays the same r\^ole that the Schr\"odinger
equation plays in the usual approach to quantum mechanics.  Since the
MDF may be interpreted as a classical
probability distribution, the Green's function connected to its
evolution equation is called the classical propagator. In 
\cite{manko1}  the classical propagator for a wide class of
Hamiltonians and its relation to the quantum propagator for the 
density matrix are  found.
This establishes an important bridge among the probability
representation of quantum mechanics and other formulations such as the
path integral approach. 

The present work deals with two different problems. On one side we 
extend the emerging new description of quantum mechanics; on the other,
we verify the formalism by applying it to a concrete, non--trivial case
of physical interest, which is  the forced parametric oscillator. A
short report on the new results which we are going to present is
already contained in \cite{noishort}. From the point of view of the
general formalism, we find new equations which connect the classical
propagator of a quantum system to its integrals of motion. Also, we
give a new definition of coherence, solely in terms of properties of
the MDF, which relates the coherent marginal distribution function to
invariants of the quantum system. In this regard let us note that the
general approach to time-dependent invariants in quantum mechanics and
their relation to the wave function were elucidated by Lewis and
Riesenfeld in  \cite{lewisprl}, while  the connection among integrals
of the motion and the quantum propagator was found in
\cite{manko75,campbell}. In \cite {urrutia} the relation of 
time-dependent invariants to the Schwinger action principle was
established, and in \cite{soliani} the relation with the N\"other theorem
was discussed. 

Let us come to the forced parametric oscillator.
This is a phenomenologically interesting model as it yields a good 
description of physical systems, like for example ions in a Paul trap. 
For such a system the symplectic tomography was already discussed in  
\cite{bregenz97} while the endoscopy scheme for measuring states was
suggested in \cite{schleich}. The trapped ion may be in different
nonclassical states like nonlinear coherent states \cite{wvogel}. A
general analysis of such states may be found in \cite{sudarmarmo}. Here 
we try a complete description of the model in the framework of the new
scheme. 

In section 1 we review the formulation of quantum mechanics in the {\it
probability representation}. We introduce the marginal distribution
function, together with its evolution equation. Then, we discuss the
relation among the classical and quantum propagators and we specialize
to the case of quadratic Hamiltonians. In section 2 we find two new 
equations for the classical propagator which is shown to be
eigenfunction of a certain time-dependent invariant. We solve these
equations for quadratic Hamiltonians and obtain the classical
propagator as a function of time-dependent invariants of the system.
This is  an important step towards the characterization of the quantum
system, as it yields both the quantum propagator for the density matrix
(which we use to test the scheme by comparing  our results to well
known results achieved by usual methods) and the time-dependence of
the MDF, once it is known  at $t=0$. In
section 3 we give a characterization of coherence directly in terms of
the MDF, and we test the new approach by
explicitly finding the coherent marginal distribution for the forced
parametric oscillator both in the new framework and by means of more
conventional methods. Finally we find the MDF 
for Fock states and we show a few significant plots. 

\section{The Probability Representation of Quantum Mechanics}
In ref. \cite{tom1} an operator $\hat X$ is discussed as a generic
linear combination of position and momentum operators
\be
\hat X=\mu \hat q + \nu\hat p  \label{x}
\ee
where $\mu, ~\nu,$ are real parameters and $\hat X$ is hermitian, hence
observable. The physical meaning of $\mu, \nu$ is that they describe an
ensemble of rotated and scaled reference frames, in classical phase
space, in which the position $X$ may be measured. In the above
mentioned paper it is shown that the quantum state of a system is 
completely determined if the classical probability distribution,
$w(X,\mu,\nu)$,  for the variable $X$, is given in an ensemble of
reference frames in the classical phase space. Such a function, also
known as the marginal distribution function, belongs to a broad
class of distributions which are determined as the Fourier transform of
a characteristic function \cite{cahill}. For the particular case  of
the variable \eqn{x}, considered in \cite{tom3}-\cite{tom2}, the scheme of 
\cite{cahill} gives 
\be
w(X,\mu, \nu)= {1\over 2\pi}\int dk e^{-ikX} <e^{ik\hat X}>~,  \label{defw}
\ee   
where $<\hat A>= \mbox{\rm Tr} (\hat\rho \hat A)$, and $\hat\rho$ is the
density operator. In \cite{cahill} it was shown that, whenever $\hat X$
is an observable, $w(X,\mu,\nu)$ is indeed a probability distribution,
as it is positive definite and satisfies the normalization condition 
\be 
\int w(X,\mu, \nu) dX =1~. \label{norm}
\ee
The definition of the MDF allows us to express it in terms of the
density matrix 
\be
w(X,\mu,\nu)={1\over 2\pi\nu} \int \rho(Z,Z') \exp \left[-i{Z-Z'\over 
\nu}\left(X-\mu{Z+Z'\over 2}\right)\right] dZ d Z'~. \label{wro} 
\ee
Recalling the relation among the Wigner function and the density 
matrix, \eqn{wro} may be rewritten as a relation among $w$ and the 
Wigner function ,
\be
w(X,\mu,\nu) =\int \exp [-ik(X- \mu q+\nu p)]W(q,p){ dk~dp~ dq\over 
(2\pi)^2}~.\label{wigner} 
\ee
  Although the general class of distribution functions of the kind
\eqn{defw} was introduced, as a function of the density matrix, already
by Cahill and Glauber in \cite{cahill}, they didn't analyze the
possibility of a new approach to quantum mechanics, in terms of such
distribution functions, mainly because the invertibility of \eqn{wro}
was not investigated. An important step in this direction is
represented by \cite{wvogel}  where Vogel and Risken have shown that
for a particular choice of the parameters $\mu$ and $\nu$ (the homodyne 
quadrature) the marginal
distribution determines completely the Wigner function via Radon
transform, namely they prove that \eqn{wigner} may be inverted for the 
 Wigner function. In the same spirit, it was shown in \cite{tom3,tom33} that
the relation among $w$ and the density matrix can be inverted for
$\rho$ yielding 
\be
\rho (X,X')={1\over 2\pi} \int w(Y,\mu,X-X') \exp \left[ 
i\left(Y-\mu{X+X'\over 2}\right)\right] d\mu d Y~. \label{row}
\ee
 and, because of such a  relation, the marginal distribution
function satisfies  an evolution equation 
\be
\del_t w + \hat O w =0~, \label{dieq}
\ee
where $\hat O$ is a finite or infinite operator polynomial in $\hat q$,
$\hat p$ (also depending on $\mu,~\nu$), determined by the Hamiltonian.
Hence, we make our previous statements more precise, by saying that the
MDF (the classical probability associated to
the random variable $X$) contains the same information on a quantum
system as the density matrix. 
The MDF may be defined through
\eqn{dieq} {\it independently} from the density matrix, hence it
represents the starting point for an alternative (but equivalent)
approach to quantum mechanics, while \eqn{dieq} may be thought of as the
analogue of the Schr\"odinger equation. Eq. \eqn{dieq} can be formally
integrated to give 
\be
w(X,\mu,\nu,t) = \int \Pi (X,\mu,\nu,0, X',\mu',\nu' ,t)  w(X',\mu',\nu',0) 
dX' d\mu' d\nu'~,\label{evopro}
\ee
where $\Pi(X,\mu,\nu,0, X',\mu',\nu',t)$ is the Green's function for the
evolution equation \eqn{dieq}. This is what we call the classical
propagator. It can be interpreted as the classical transition
probability density from an initial position $X'$ in the ensemble of
reference frames of the classical phase space, to the position $X$
\cite{tom33}. 

Let us now elucidate the connection of the classical propagator with
the quantum propagator (Green function) for the density matrix
$\rho(X,X',t)$. Details may be found in \cite{manko1,new}. For a
pure state with wave function $\Psi(X,t)$, we have 
\be 
\rho(X,X',t)= \Psi(X,t) \Psi^*(X',t).
\ee
Since the wave function at time $t$ is connected to the one at initial
time by the Green's function of the Schr\"odinger equation $G(X, X', t)$
\be
\Psi(X,t) = \int \Psi(X',0) G(X, X', t) d X'
\ee
we have for the density matrix
\be
\rho(X,X',t)= \int K(X,X',Y,Y',t)\rho(Y,Y',0) dY dY'
\ee
with
\be
K(X,X',Y,Y',t)=G(X, Y, t)G^*(X',Y', t). \label{qprop}
\ee
The function $K(X,X',Y,Y',t)$ is what is called the quantum propagator
for the density matrix. Using the relation between the density matrix
and the MDF \eqn{row} we finally find 
\beqa
K(X,X',Z,Z',t)&=&{1\over (2\pi)^2} \cr
&\times& \int {1\over\nu'} \exp\left\{i\left(Y-\mu{X+X'\over 2}\right)
-i{Z-Z'\over \nu'}Y' +i {Z^2+Z'^2\over 2\nu'}\mu'\right\}\cr
&\times& \Pi(Y,\mu, X-X',0,Y',\mu',\nu',t) d\mu d\mu' dY dY' d\nu'~. 
\label{qprop2}
\eeqa
Then, once the classical propagator is known, the quantum propagator
for the density matrix can be found. In the next sections, after finding 
the classical propagator, we will give an
explicit example of this kind of calculation, for the driven parametric
oscillator. We will also compare our results to analogous calculations
obtained with the method of path integrals \cite{feynman}. 

\section{The classical propagator}
\setcounter{equation}{0}
In this section we address the problem of finding the classical 
propagator, in the framework of the time-dependent invariants method. 
We consider Hamiltonians  of the form 
\be
\hat H = \frac{\hat p^2}{2} + V(\hat q) \label{quad}
\ee
where $V(q)$ is a generic potential energy. In this case 
the evolution equation for the MDF, Eq. \eqn{dieq}, takes
the form \cite{tom3,tom33,manko1} 
\be
\dot w -\mu {\del\over \del\nu} w -i \left[V\left(\frac{-1}{\del/\del 
x}\frac{\del}{\del \mu} -i\frac{\nu}{2}\frac{\del}{\del x}\right) -
V\left(\frac{-1}{\del/\del 
x}\frac{\del}{\del \mu} +i\frac{\nu}{2}\frac{\del}{\del 
x}\right)\right]w=0. \label{fokker}
\ee 
Restoring the proper units, the Planck constant $\hbar$ will appear 
in Eq. \eqn{fokker} so that the equation, even if
classical-like, gives a quantum description of the system evolution,
so replacing the Schr\"odinger equation in our scheme. It can be shown  
\cite{new} that in classical statistical mechanics the distribution 
$w(X,\mu,\nu)$ may be also introduced and the classical Boltzmann 
equation can be rewritten for this distribution. Then the classical
limit of \eqn{fokker} is the Boltzmann equation. The classical propagator
obeys an evolution equation which follows from \eqn{fokker} 
\beqa
\frac{\del \Pi}{\del t_2}  -\mu {\del\over \del\nu} \Pi -i 
\left[V\left(\frac{-1}{\del/\del 
x}\frac{\del}{\del \mu} -i\frac{\nu}{2}\frac{\del}{\del x}\right) -
V\left(\frac{-1}{\del/\del 
x}\frac{\del}{\del \mu} +i\frac{\nu}{2}\frac{\del}{\del 
x}\right)\right]\Pi \cr
=\delta(t_2-t_1)\delta(X'-X)\delta(\mu'-\mu)
\delta(\nu'-\nu), \label{evpi}
\eeqa 
with initial condition
\be
\Pi(X,\mu,\nu,0, X',\mu',\nu' ,0)=\delta(X'-X)\delta(\mu'-\mu)
\delta(\nu'-\nu) ~.
\ee
To be definite, let us consider the Hamiltonian  describing the forced
parametric oscillator. This is of the form described by \eqn{quad}, and it
includes many other quadratic Hamiltonians as limiting cases. The potential,
$V(q)$, is given by 
\be
V(q)  = {\omega^2(t) \over 2} q^2 - f(t)q, \label{potential}
\ee 
that is 
\be
H={p^2\over 2} + {\omega^2(t) \over 2} q^2 - f(t)q~.
\ee
We leave $\omega$ free to be real or imaginary, to allow the
description of repulsive oscillators as well. The evolution equation
for the marginal distribution $w$, is obtained by the general
expression \eqn{fokker}, replacing the potential by \eqn{potential}. We
have 
\be
\dot w - \mu {\del \over \del \nu} w + \left(\omega^2(t) \nu  
{\del \over \del \mu} + f(t) \nu {\del \over \del X}\right) w =0~.
\label{evosc}
\ee
The equation for the propagator may be obtained analogously.

The solution to \eqn{evpi} was 
previously argued to be expressible in terms of the integrals of the
motion of the system \cite{new}. We show here a derivation of  
this result, which is essentially an extension of previous techniques 
\cite{lewisprl,manko75,campbell,soliani,urrutia}. In the mentioned 
papers the time-dependent invariants of a given system were shown to 
be in connection with the wave function and with the Green function 
of the Schr\"odinger equation. 
The integrals of the motion, $I(t)$ are defined by the equation
\be
{\del \over \del t} I(t)  + i [H, I(t)] =0~, \label{ieq}
\ee
where $H$ is the Hamiltonian of the system.
We may think of time-dependent invariants as the evolution in time of 
the initial coordinates and momenta, $q$ and $p$. Then we express them 
as 
\be
I(t)= \Lambda(t) Q + \Delta(t) \label{imo}
\ee
with $Q=(p,q)$ and $\Lambda, \Delta$ to be determined. 
For the oscillator any integral of the motion may be expressed
as a function of the two operators $A(t)=UaU^{-1}$ and
$A^{\dag}(t)=Ua^\dag U^{-1}$, where $U$ is the evolution operator
obtained by the Schr\"odinger equation. Of course, they are integrals of 
the motion in their turn. We have 
\beqa
A(t)&=& {i\over \sqrt 2} (\epsilon(t) p - \dot \epsilon (t) q) + \beta (t)
 \label{aa} \\
A^{\dag}(t)&=& 
-{i\over \sqrt 2} (\epsilon^*(t) p - \dot \epsilon^* (t) q) + \beta^* (t),
\label{adef}
\eeqa
with $\epsilon$ satisfying
\beqa
\ddot\epsilon + \omega^2(t) \epsilon&=&0  \label{epsi}\\
\dot\epsilon \epsilon^*- \dot\epsilon^*\epsilon&=&2i
\eeqa
and initial conditions $\epsilon(0)=1$, $\dot\epsilon(0)=i$.
It can be checked that
\be
[A, A^{\dag}]=1
\ee
The function $\beta(t)$ is determined by consistency with \eqn{ieq} to be 
\be
\beta(t)= -{i\over \sqrt 2} \int_0^t dt'~ \epsilon(t') f(t')~. 
\label{beta}
\ee
The integral of the motion 
\be
I(t)= \pmatrix{I_p \cr I_q}= \pmatrix {\Lambda_{11} & \Lambda_{12} \cr
\Lambda_{21}& \Lambda_{22} } \pmatrix{ p \cr q} + \pmatrix{ \Delta_1 \cr 
\Delta_2} \label{it}
\ee
can be given in terms of   $A$ and $A^{\dag}$ as
\beqa 
I_p &=& {A-A^{\dag}\over \sqrt 2 i}\cr
I_q &=& {A+A^{\dag}\over \sqrt 2 }~.    \label{ii}
\eeqa
The matrix Lambda and the vector $\Delta$ are then determined by
comparing \eqn{it} with \eqn{ii}. We obtain 
\be
\Lambda={1\over 2}\pmatrix{\epsilon+\epsilon^* & -(\dot\epsilon+\dot
\epsilon^*) \cr 
i(\epsilon-\epsilon^*) & -i(\dot\epsilon-\dot\epsilon^*) 
}\label{lambda}
\ee
and
\be
\Delta={1\over\sqrt 2}\pmatrix{i(\beta-\beta^*)\cr \beta+\beta^*}. 
\label{gdelta}
\ee
In \cite{manko75,campbell} it was shown that the Green function 
$G(q,q',t)$ is a solution of the system
\beqa
I_q G(q,q',t)&=&{\hat q'}G(q,q',t) \label{qgreen}\\
I_p G(q,q',t)&=& -{\hat p'}G(q,q',t) , \label{pgreen}
\eeqa
where the first equation means that the Green's function is an
eigenfunction of the invariant $I_q$ at each value of $t$, with
eigenvalue the initial position, $q'$, of the system. 
These results were derived by applying the evolution operator 
$U(t)$ to the identities 
\beqa
{\hat q} \delta (q-q')&=&{\hat q}'\delta(q-q') \cr
{\hat p} \delta (q-q')&=& -{\hat p'} \delta(q-q')~,
\eeqa
where $G(q,q',0)=\delta(q-q')$. Equations \eqn{qgreen} and \eqn{pgreen} may
be trivially generalized to equations for the quantum propagator of the
density matrix $K(X,X',Y,Y',t)$. Thus in principle we can find analogous
relations for the classical propagator $\Pi$, inverting the relation
\eqn{qprop2}. This procedure, although mathematically well posed is in 
practice difficult to pursue. We will use instead a formal procedure 
which goes along the same lines of the derivation for the Green's function. 

Assuming the existence of an evolution operator, $\tilde U$, for the
equation \eqn{evosc}, and recalling that $\Pi(X,\mu,\nu,0,  X',\mu',\nu'
,0)= \delta(X-X')\delta(\mu-\mu')\delta(\nu-\nu')$, we find 
\beqa
{\tilde I}_q \Pi(X, \mu,\nu, 0, X',\mu',\nu',t) &=& { \tilde q'}
\Pi(X, \mu,\nu,0, X',\mu',\nu',t)   \label{xclprop}\\
{\tilde I}_p \Pi(X, \mu,\nu,0, X',\mu',\nu',t) &=& - {\tilde p'} 
\Pi(X, \mu,\nu,0, X',\mu',\nu',t)   
 \label{pclprop}~.
\eeqa
Let us explain the notation. We mean with $\tilde q'$ and $\tilde p'$ the 
operators which represent the action which is induced on $w(X, \mu, \nu, t)$ 
when acting on the density matrix with $\hat q'$ and $\hat p'$, due to 
\eqn{row}. Analogously ${\tilde I}_q$ and  ${\tilde I}_p$ represent the 
action which is induced on $w(X,\mu,\nu,t)$ when acting on the density 
matrix with $\hat I_q$ and $\hat I_p$. The operators ${\tilde I}_q$ and  
${\tilde I}_p$ are formally connected to  $\tilde q$ and $\tilde p$ 
through the evolution operator $\tilde U$.
They are given by
\bean
\tilde q'&=& -\left(\frac{\partial}{\partial X'}\right)^{-1}
\frac{\partial}{\partial\mu'}-i\frac{\nu'}{2}\frac{\partial}{\partial X'} \\
\tilde p'&=&  -\left(\frac{\partial}{\partial X'}\right)^{-1}
\frac{\partial}{\partial\nu'}+i\frac{\mu'}{2}\frac{\partial}{\partial X'} \\
{\tilde I}_q&=& \frac{1}{4}\left[\mu(\epsilon-\epsilon^*)+\nu(\dot\epsilon-
\dot\epsilon^*) \right]\frac{\partial}{\partial X}+\frac{i}{2}\left[
(\dot\epsilon-\dot\epsilon^*) \frac{\partial}{\partial\mu} - (\epsilon-\epsilon^*) 
\frac{\partial}{\partial\nu}\right]\left(\frac{\partial}{\partial X}\right)^{-1} 
+\frac{\beta+\beta^*}{\sqrt{2}}\\
{\tilde I}_p&=& -\frac{i}{4}\left[\mu(\epsilon+\epsilon^*)+\nu(\dot\epsilon+
\dot\epsilon^*) \right]\frac{\partial}{\partial X}+\frac{1}{2}\left[
(\dot\epsilon+\dot\epsilon^*) \frac{\partial}{\partial\mu} - (\epsilon + \epsilon^*) 
\frac{\partial}{\partial\nu}\right]\left(\frac{\partial}{\partial X}\right)^{-1} 
+\frac{\beta-\beta^*}{\sqrt{2}}~.
\eean
The solution to these equations can be verified to be 
\be
\Pi(X,\mu,\nu,0, X',\mu',\nu' ,t)=\delta(X-X'+{\cal N}\Lambda^{-1}\Delta) 
\delta({\cal N}'-{\cal N}\Lambda^{-1}) \label{clprop}
\ee
where ${\cal N},~{\cal N}'$ are vectors,  ${\cal N}=(\nu, \mu),~~~ {\cal 
N}'=(\nu', \mu')$. 

Substituting \eqn{lambda} and \eqn{gdelta} into \eqn{clprop} we 
get the classical propagator to be 
\beqa
\Pi(X,\mu,\nu,0, X',\mu',\nu' ,t)&=&\delta\left(X-X'+ 
\frac{\nu(\dot\epsilon^*\beta+\dot\epsilon\beta^*)+
\mu(\epsilon^*\beta+\epsilon\beta^*)}{\sqrt 2}\right)\label{cprop2} \\
&\times & \delta\left(\nu' -i\frac{\nu(\dot\epsilon^*-\dot\epsilon)+
\mu(\epsilon^*-\epsilon)}{2}\right)
 \delta\left(\mu' -\frac{\nu(\dot\epsilon^*+\dot\epsilon)+
\mu(\epsilon^*+\epsilon)}{2}\right)~.\nonumber 
\eeqa 
Now, we can replace this expression into \eqn{qprop2} and we obtain the
quantum propagator for the density matrix of the driven parametric
oscillator. Of course, the integral cannot be performed unless we
assign the explicit dependence of the frequency, $\omega(t)$, so that
we can solve \eqn{epsi} for $\epsilon(t)$. Here we  assume $\omega=$
constant. We have then 
\be
\epsilon(t)= e^{it}~,
\ee  
and the classical propagator \eqn{cprop2} assumes the form
\beqa
\Pi(X,\mu,\nu, 0,X',\mu',\nu' ,t)&=&\delta\left(X-X'+ 
\frac{\beta e^{-it} (\mu-i\nu)+\beta^* e^{it} (\mu+i\nu)}{\sqrt 2}\right)\cr
&\times& \delta\left(\nu' -(\nu\cos t +\mu\sin t)\right)
\times \delta\left(\mu' +(\nu\sin t-\mu\cos t) \right)~.
\label{cprop3}
\eeqa 
Substituting into \eqn{qprop2}, this yields the quantum propagator 
{
\footnotesize
\beqa
K(X,X',Z,Z',t)&=&{1\over (2\pi)^2} \int {1\over\nu'} 
\exp\left\{i\left(Y-\mu{X+X'\over 2}\right)
-i{Z-Z'\over \nu'}Y' +i {Z^2+Z'^2\over 2\nu'}\mu'\right\}\cr
&\times& 
\delta\left(Y-Y'+ 
\frac{\beta e^{-it} (\mu -i(X-X'))+\beta^* e^{it} (\mu +i(X-X'))}{\sqrt 2}
\right)\cr
&\times& \delta\left(\nu' -((X-X')\cos t +\mu\sin t)\right)
\times \delta\left(\mu' +((X-X')\sin t-\mu\cos t) \right)
d\mu d\mu' dY dY' d\nu'~. 
\label{qprop3}
\eeqa
}
Performing the integration we get
{
\small
\beqa
K(X,X',Z,Z',t)&=&{1\over 2\pi \sin t} 
\exp\left\{ {-i\over \sqrt{2}} \left[\beta e^{-it}\left(-i(X-X') +  
{Z-Z'\over \sin t} - {(X-X')\cos t \over \sin t} \right)\right]\right\}  \cr
&\times&\exp\left\{ {-i\over \sqrt{2}} \left[\beta^* e^{it}\left(i(X-X') 
+  {Z-Z'\over\sin t} - {(X-X')\cos t \over \sin t} \right)\right]\right\} \cr
&\times& \exp\left\{ -i{(X+X')\over 2} 
\left[{Z-Z'\over \sin t} - {(X-X')\cos t \over \sin t} 
\right]\right\}\label{qprop4}\\
 &\times& \exp\left\{
{i\over 2}(Z+Z')\left[ -(X-X')\sin t + {(Z-Z')\cos t\over \sin t} + 
{(X-X')\cos^2 t \over \sin t}\right]\right\}, \nn
\eeqa
}
where we have used
$$
\int dY' \exp\left\{iY'\left[1- {(Z-Z')\over \mu \sin t +(X-X')
\cos t}\right]\right\}= 2\pi \delta\left(1- {(Z-Z')\over \mu \sin t 
+(X-X')\cos t}\right)
$$
and 
$$
\delta\left(f(\mu)\right) = {1\over |f'(\mu_0)|} \delta (\mu-\mu_0),
$$
with $\mu_0$ s. t. $f(\mu_0)=0$.

In view of interpreting the quantum propagator for the density matrix
as the product of quantum propagators for the wave function, as in
\eqn{qprop}, we separate the primed and unprimed variables in
\eqn{qprop4}, obtaining 
\beqa
K(X,Z,X',Z',t)&=& {1\over 2\pi\sin t} 
\exp\left\{ {i\over 2\sin t} [(X^2+Z^2) \cos t - 2XZ]\right\}\cr
&\times&\exp\left\{
{-i\over \sqrt{2}\sin t} [Z(\beta e^{-it} + \beta^* e^{it}) - 
X(\beta+\beta^*)]\right\}\cr
&\times& \exp\left\{ -{i\over 2\sin t} [(X'^2+Z'^2) \cos t - 2X'Z']\right\}\cr
&\times&\left\{
{i\over \sqrt{2}\sin t} [Z'(\beta^* e^{it} + \beta e^{-it}) - 
X'(\beta^*+\beta)]\right\}~.
\eeqa
This expression can be further simplified using the explicit form of 
the shift, $\beta$, as given in \eqn{beta}. We finally get
\beqa
K(X,Z,X',Z',t)&=& {1\over 2\pi\sin t} 
\exp\left\{ {i\over 2\sin t} [(X^2+Z^2) \cos t - 2XZ]\right\} \cr
&\times& \exp\left\{
2Z \int_0^t f(t') \sin(t-t') dt' +2X \int_0^t f(t') \sin t' dt'
\right\}\cr
&\times& \exp\left\{ -{i\over 2\sin t} [(X'^2+Z'^2) \cos t - 2X'Z'] \right \}
\cr
&\times& \exp\left\{
2Z' \int_0^t f(t') \sin(t-t') dt' +2X' \int_0^t f(t') \sin t' dt'
\right\}~. \label{qprop5}
\eeqa 
Now we can read off the quantum propagator for the wave function. It is
worth noting that it can be determined only  up to a phase factor,
independent on phase space, but possibly dependent on time, as can be
argued from \eqn{qprop}. We have 
{
\small
\beqa
 G(X,Z,t)&=& e^{i F(t)} {1\over \sqrt{2\pi\sin t}} 
\exp\left\{ {i\over 2\sin t} [(X^2+Z^2) \cos t - 2XZ]\right\} \cr
&\times& \exp\left\{{i\over 2\sin t} \left[
2Z \int_0^t f(t') \sin(t-t') dt' +2X \int_0^t f(t') \sin t' dt'
\right]\right\}, \label{fey}
\eeqa
}
where $F(t)$ is an unknown, real function. This is the expected result,
already obtained in the literature with other techniques (cfr. for
example \cite{feynman}, where the quantum propagator  is obtained with
path integral methods). Thus, we have checked in a specific example
that the probability representation gives equivalent predictions to
the standard formulations of quantum mechanics. 

The quantum propagator for the simple harmonic oscillator
($\omega=$constant, $\beta=0$) is easily obtained to be 
\beqa
K(X,Z,X',Z',t) &=& {1\over 2\pi\sin t}
\times \exp\left\{ {i\over 2\sin t} [(X^2+Z^2) \cos t - 2XZ] 
\right\}\cr
&\times& \exp\left\{ -{i\over 2\sin t} [(X'^2+Z'^2) \cos t - 2X'Z'] 
\right\},
\eeqa 
yielding
\be
G(X,Z,t)=  e^{iF(t)} {1\over \sqrt{2\pi\sin t}} 
\times \exp\left\{ {i\over 2\sin t} [(X^2+Z^2) \cos t - 2XZ] 
\right\}~.
\ee
The quantum propagator for the free motion, already calculated with the
illustrated techniques in \cite{manko1,new}, may be recovered in the
limit $\beta=0, \omega=0$ (which  corresponds to $ ~ \sin t \rightarrow
t,~\cos t \rightarrow 1$). We get 
\be
K(X,X', Z, Z',t)= {1\over 2\pi t} \exp\left\{ {i\over 2t} [-(X-Z)^2 + 
(X'-Z')^2]\right\}
\ee 
yielding
\be
G(X,Z,t)= e^{iF(t)} {1\over \sqrt{2\pi t}} 
\times \exp\left\{ {i\over 2t} [(X-Z)^2] 
\right\}~. 
\ee

\section{The marginal distribution for coherent states}
\setcounter{equation}{0}
In this section we find the marginal distribution function for coherent
states, using three techniques. First we give two derivations which are
based on a reformulation of the notion of coherence in terms of the
distribution function itself. These derivations are particularly
relevant from a conceptual point of view  giving another example of how
common concepts of quantum mechanics are treated in the new approach
without recursion to the wave function. Then we use the relation among
the distribution function and the density matrix given by \eqn{row}. In
this approach, the notion of coherence is defined in the conventional
manner, through the wave functions, by requiring that they be solutions
of the Schr\"odinger equation with initial condition 
\be 
a \psi_\alpha=\alpha \psi_\alpha \label{ico}
\ee
(with $a, a^\dag$ usual annihilation and creation operators). As known
 the two equations (the Schr\"odinger equation and the initial
condition) may be put together, yielding 
\be 
A(t)\psi_\alpha = \alpha \psi_\alpha~. \label{coco} 
\ee 
where the operator $A$, defined in the previous section  by \eqn{adef},
represents the time evolution of $a$. Once we find the solution of
\eqn{coco}, we can write the density matrix associated to coherent
states, and, consequently, the MDF, through \eqn{row}. 

\subsection{Probability Representation Approach}

Let us describe the new approach, in the spirit of the
probability representation of quantum mechanics. The first problem we
are faced with, is to define the notion of coherence of a quantum
state, independently from wave functions. We want coherent states to be
defined as those states whose MDF satisfies
the Fokker-Planck equation \eqn{fokker}, with an initial condition
which ``translates'' \eqn{ico} into an equation  for $w$. Applying the
annihilation operator $a$ to the density matrix, and using \eqn{ico} we
get 
\be 
a\rho(X,X')= \alpha \rho(X,X')~ \label {aro}
\ee
(acting on the right with $a^\dag$ we get an equivalent equation, with
$\alpha^*$ instead than $\alpha$). Now, recalling that $a={1\over
\sqrt{2}} \left(X+{\del\over\del X}\right)$ and using the expression of 
the density matrix in terms of the MDF, \eqn{row}, Eq. \eqn{aro} yields
\be
{1\over \sqrt{2}} \left\{{\mu+i\nu \over 2} {\del\over\del X} + 
\left( {\del \over \del \mu} + i {\del \over \del \nu}  
\right){\del\over\del X}^{-1}\right\} w_{\alpha} (X,\mu,\nu) = \alpha 
w_{\alpha} (X,\mu,\nu), \label{cow}
\ee
where we have used the correspondence \cite{new} 
\beqa
{\del\over \del X} \rho(X,X') &\rightarrow& \left[{1\over 2} \mu {\del\over
\del X} - i \left( {\del\over \del X}\right)^{-1} {\del\over
\del\nu}\right]w(X,\mu,\nu)  \cr 
X\rho(X,X') &\rightarrow& -\left[\left( {\del\over \del X}\right)^{-1} 
{\del\over \del\mu} +{i\over 2} \nu {\del\over \del X} \right] 
w(X,\mu,\nu) ~. \label{corri}
\eeqa  
Eq. \eqn{cow} may be rewritten in the compact form
\be
\tilde a w_\alpha(X,\mu,\nu)= \alpha w_\alpha(X,\mu,\nu),
\ee
where the operator 
\be
\tilde a = {1\over \sqrt{2}} \left\{{\mu+i\nu \over 2} {\del\over\del X} + 
\left( {\del \over \del \mu} + i {\del \over \del \nu}  
\right){\del\over\del X}^{-1}\right\}~
\ee
translates the action of $a$ on the density matrix into an action on 
$w$.
Hence, we may reformulate the problem of finding the MDF for coherent
states entirely in the language of classical probability, the initial
condition \eqn{ico} being replaced by \eqn{cow}. The solution of
\eqn{cow} represents $w_\alpha$ at $t=0$. We get the solution at a
generic value of $t$ just as we would do for wave functions, namely, by
applying the propagator for the corresponding evolution equation. For
wave functions this is the Green function of the Schr\"odinger
equation, while for the MDF it is the classical propagator which we
found in the previous section (the Green function of the Fokker-Planck
equation). To solve \eqn{cow} we make a Fourier transform of the
marginal distribution function 
\be
w_\alpha (X,\mu,\nu)=\int w_k( \alpha, \mu, \nu) e^{ikX} dX
\ee
and we get 
\be
\left[ik {\mu+i\nu\over 2} - {1\over ik} \left({\del\over \del\mu} + i 
{\del\over\del\nu}\right)\right]w_k=\sqrt{2}\alpha w_k~.
\ee 
This equation is of the form 
\be
\left[\left({i\over 2} y +i{\del\over \del y}\right) -\left( 
{z\over 2} -{\del\over\del z}\right)\right]w_k=\sqrt{2}\alpha w_k~,
\label{afou}
\ee 
with $ y= k\mu$, $z=k\nu$. The solution of such an equation is known 
to be a Gaussian 
\be
w_k\sim \exp{\left(c y^2+dz^2 +hyz+ey+gz\right)}~. \label{gau}
\ee
It turns out that, to determine the coefficients, we also need the
complex conjugate of \eqn{afou},
\be
\left[\left({i\over 2} y +i{\del\over \del y}\right) +\left( 
{z\over 2} -{\del\over\del z}\right)\right]w_k=\sqrt{2}\alpha^* w_k.~
\label{afou2}
\ee 
where we have used
\be
w_k^*(y,z)=w_{-k}(y,z)=w_k (-y,-z) 
\ee 
(remember that $w_\alpha$ is real).
Hence, we determine $c, d, e, g, h$  by consistency to be 
\be
c= d= -{1\over 4}~,~~~~e={\alpha^*-\alpha\over \sqrt{2}}~,~~~~
g=-i{\alpha^*+\alpha\over \sqrt{2}}~,~~~h=0.
\ee
Restoring the original notation
we have
\be
w_k(\alpha,\mu,\nu)= N\exp\left\{-k^2{(\mu^2+ \nu^2) \over 4} +
{\alpha^*-\alpha\over \sqrt{2}} k\nu  -i {\alpha^*-\alpha\over \sqrt{2}} 
k\mu \right\}, \label{ftr}
\ee
where $N$ is a normalization factor. Taking the inverse Fourier transform of 
\eqn{ftr} we finally get
\be
w_\alpha(X,\mu,\nu)=   N{1\over\sqrt{\pi(\mu^2+\nu^2)}} 
\exp\left\{-{(X+i{\alpha-\alpha^*\over\sqrt{2}}\nu - {\alpha+ \alpha^*\over 
\sqrt{2}}\mu)^2\over \mu^2+\nu^2}\right\} ~.\label{zerow}
\ee
The last step is to find the time dependence of the coherent marginal 
distribution function. This can be achieved by substituting the 
solution at $t=0$, \eqn{zerow}, into Eq. \eqn{evopro}, where the propagator 
is given by \eqn{cprop2}. We have
\beqa
w_\alpha(X,\mu,\nu,t) &=& 
N \int \Pi (X,\mu,\nu, 0,X',\mu',\nu' ,t)  w_\alpha (X',\mu',\nu',0) 
dX' d\mu' d\nu'\cr
&=& N \int \delta\left(X-X'+ 
\frac{\nu(\dot\epsilon^*\beta+\dot\epsilon\beta^*)+
\mu(\epsilon^*\beta+\epsilon\beta^*)}{\sqrt 2}\right)\cr
&\times& \delta\left(\nu' -i\frac{\nu(\dot\epsilon^*-\dot\epsilon)+
\mu(\epsilon^*-\epsilon)}{2}\right)
 \delta\left(\mu' -\frac{\nu(\dot\epsilon^*+\dot\epsilon)+
\mu(\epsilon^*+\epsilon)}{2}\right) \cr
&\times& \sqrt{{1\over\pi(\mu'^2+\nu'^2)}} 
\exp\left\{-{(X'+i{\alpha-\alpha^*\over\sqrt{2}}\nu' - {\alpha+ \alpha^*\over
\sqrt{2}}\mu')^2\over \mu'^2+\nu'^2}\right\} dX' d\mu' d\nu' \nonumber
\eeqa
which, performing the integration, yield
\be
w_\alpha(X,\mu,\nu,t) = 
N {1\over \sqrt{\pi
|{\dot\epsilon\nu} + \epsilon\mu|^2} } 
\exp \left\{ -{1\over |{\dot\epsilon\nu} + \epsilon\mu|^2} 
\left[X -  {\gamma(\epsilon^*\mu+\nu\dot\epsilon^*) +  
\gamma^* (\epsilon\mu + \dot\epsilon\nu)\over \sqrt{2}}\right]^2\right\}
\label{wal2}~.\nonumber
\ee
The normalization factor, $N$, is equal to 1, due to the normalization
condition \eqn{norm}. 

It is worth noting that, whenever the time dependence is not too
complicate, the previous calculation to get the time--dependent 
marginal distribution function, which consists of two steps, may be 
replaced by directly solving the time--dependent equation 
\be
\tilde A(t) w_\alpha(X,\mu,\nu,t)= \alpha w_\alpha(X,\mu,\nu,t)~ 
\label{ticoco}
\ee
which is obtained in the same way as \eqn{coco} as a straightforward
extension of the procedure illustrated for the time independent case.
The operator $\tilde A(t)$, which we will find explicitly below is
obtained from the action of $A(t)$ on the density matrix when we
express $\rho(X,X',t)$ as a function of $w_\alpha$. In our situation
\eqn{ticoco} turns out to be solvable using the same techniques which
we used to solve the time independent analogue. This derivation is
completely equivalent to the previous one. We choose to present both of
them, because one requires the explicit use of the classical
propagator, emphasizing the great content of information contained in
it, while the other shows how simple calculations can be in the
framework of the probability representation of quantum mechanics. 
  
Since the operator  $A(t)$ given by \eqn{aa} may be rewritten in 
coordinate representation (of phase space) as
\be
A(t)= {1\over \sqrt{2}}\left(\epsilon(t) {\del\over \del X} 
-i\dot\epsilon(t) X\right) + \beta(t),
\ee
we can use the correspondence \eqn{corri} to write \eqn{ticoco} as
\be
{1\over \sqrt{2}} \left\{{\epsilon \mu+\dot\epsilon\nu \over 2} 
{\del\over\del X} + i
\left( \dot\epsilon{\del \over \del \mu} -\epsilon {\del \over \del \nu}  
\right)\left( {\del\over\del X}\right)^{-1} \right\} 
w_{\alpha} (X,\mu,\nu) = \gamma 
w_{\alpha} (X,\mu,\nu)~, \label{ticow}
\ee
where $\gamma=\alpha-\beta$.
As before, this can be easily solved performing a Fourier transform 
of the MDF. We get
\be
\left[\left({i\over 2} \epsilon y +\dot\epsilon{\del\over \del y}\right) 
+i\left( {\dot\epsilon z\over 2} -\epsilon {\del\over\del z}\right)\right]w_k
=\sqrt{2}\gamma w_k~,
\label{afouu}
\ee 
with $ y= k\mu$, $z=k\nu$. The solution is a Gaussian of the form 
\eqn{gau}. To determine the coefficients we proceed as before. We 
replace \eqn{gau} into \eqn{afouu}. We get a consistency equation for 
the coefficients, which doesn't determine them completely. Hence, we 
take the complex conjugate of \eqn{afouu}, and we use the fact that 
$w^*_k=w_{-k}= w_k(-y,-z)$. We obtain
\beqa
\left({i\epsilon \over 2} +2\dot\epsilon d -\epsilon h\right) y
+  \left({i\dot\epsilon \over 2}  -2\epsilon c +\dot\epsilon h\right)z 
+\dot\epsilon e - \epsilon f = \sqrt{2}\gamma\cr
\left({i\epsilon^* \over 2} -2\dot\epsilon^* d +\epsilon^* h\right) y
+  \left({i\dot\epsilon^* \over 2}  +2\epsilon^* c -\dot\epsilon^* h\right)z 
-\dot\epsilon^* e + \epsilon^* f = \sqrt{2}\gamma^*,
\eeqa
which yields
\be
c=-{|\dot\epsilon|^2\over 4}~,~~~d=-{|\epsilon|^2\over 4}~,
e={\gamma\epsilon^*+\gamma^*\epsilon\over i\sqrt{2}}~,~~~
f={\gamma\dot\epsilon^*+\gamma^*\dot\epsilon\over i\sqrt{2}}~,~~~
h=-{\epsilon\dot\epsilon^*+ \epsilon^*\dot\epsilon\over 4}~.
\ee
Restoring the original notation we thus get
\beqa
w_k(\alpha, \mu,\nu,t)  & = & 
N \exp\left\{-{k^2\over 4} \left[\mu^2|\epsilon|^2 + \nu^2 |\dot\epsilon|^2  +
+(\dot\epsilon\epsilon^*+ \dot\epsilon^*\epsilon)\mu\nu\right] \right\} \cr
& \times & \left\{-ik             
\left( \mu{\gamma\epsilon^*+\gamma^*\epsilon\over\sqrt{2}} + 
\nu {\gamma\dot\epsilon^*+\gamma^*\dot\epsilon\over\sqrt{2}}\right) \right\}
\label{ftra}
\eeqa
Taking the Fourier transform of this expression, it is immediately 
verified that it is identical to \eqn{wal2}, as expected.

\subsection{The Schr\"odinger Approach}
Let us come to the more conventional point of view. We first find the
coherent wave functions and the corresponding density matrix, and then
obtain the coherent marginal distribution function through the Eq. \eqn{wro}.
The solution to Eq. \eqn{coco} is of the form 
\be
\psi_\alpha (t)= C(t) \exp\left(i\frac{x^2}{2} 
\frac{\dot\epsilon}{\epsilon} + 
\frac{\sqrt{2}}{\epsilon}(\alpha-\beta)x\right) ~. \label{psi}
\ee
We determine the overall factor, $ C(t)$, up to a phase factor, by
imposing the wave function to be normalized to 1. We pose $C(t)= D(t)
\exp(i\phi)$, we have then 
\be
1= \int_{-\infty}^\infty |\psi_\alpha|^2 dx = D(t)^2 \sqrt{\pi \epsilon 
\epsilon^*} \exp\left\{ {1 \over 2\epsilon\epsilon^*} \left[(\alpha- 
\beta) \epsilon^* + (\alpha^*-\beta^*)\epsilon\right]^2\right\},
\ee
from which we get
\be
\psi_\alpha (t) = \frac{e^{i\phi}}{ (\pi\epsilon\epsilon^*)^{1/4}}
 \exp\left\{-{1\over 4\epsilon\epsilon^*} 
\left[(\alpha- 
\beta) \epsilon^* + (\alpha^*-\beta^*)\epsilon\right]^2\right\}
   \exp\left(i\frac{x^2}{2} 
\frac{\dot\epsilon}{\epsilon} + 
\frac{\sqrt{2}}{\epsilon}(\alpha-\beta)x\right)~.  \label{psii}
\ee
We now replace the solution we found in \eqn{row}, recalling that
\be
\rho_\alpha(Z,Z')= \psi_\alpha(Z)\psi^*_\alpha(Z')~.
\ee
After some algebra we get
\beqa 
w_\alpha(X,\mu,\nu)&=&{1\over 2\pi\nu\sqrt{\pi\epsilon\epsilon^*}} 
\exp\left[{-1\over 2\epsilon\epsilon^*} (\gamma \epsilon^*+ \gamma^* 
\epsilon)^2\right] \times \cr
&&\int dZ \exp\left[{i\over 2} \left({\dot\epsilon\over\epsilon} + 
{\mu\over\nu}\right)Z^2 + \left( \sqrt{2} {\gamma\over\epsilon} - {i\over 
\nu} X \right) Z\right]\times \cr
&& \int dZ' \exp\left[-{i\over 2} \left({\dot\epsilon^*\over\epsilon^*} + 
{\mu\over\nu}\right)Z'^2 + \left( \sqrt{2} {\gamma^*\over\epsilon^*} + {i\over 
\nu} X \right) Z'\right],
\eeqa
where we have posed $\gamma = \alpha-\beta$. The Z, Z' dependence is
completely factorized and the two integrals can be put into Gaussian
form so that 
\beqa
w_\alpha(X,\mu,\nu)&=& {1\over \nu\sqrt{\pi\epsilon\epsilon^*}
| {\dot\epsilon\over\epsilon} + {\mu\over\nu} | } 
\exp \left\{ -{1\over 2 \epsilon\epsilon^*} 
(\gamma\epsilon^*+\gamma^*\epsilon)^2 \right\} \times \cr
&\times & \exp \left\{ {i\over 2
\left( {\dot\epsilon\over\epsilon} + {\mu\over\nu}\right)} 
\left({2\gamma^2\over\epsilon^2} -{X^2\over \nu^2}- 
{2\sqrt{2}i\over \nu}{\gamma\over\epsilon} X\right)\right\}\times\cr
&\times &\exp\left\{-{i\over 2 \left( 
{\dot\epsilon\over\epsilon} + 
{\mu\over\nu}\right)^*}\left({2\gamma*^2\over\epsilon*^2} -{X^2\over 
\nu^2}+ 
{2\sqrt{2}i\over \nu}{\gamma^*\over\epsilon^*} X\right) \right\}~.
\eeqa
This rather complicated expression becomes simpler if we recognize the 
exponential factor as a square of three terms:
\be
w_\alpha(X,\mu,\nu)= {1\over \nu\sqrt{\pi\epsilon\epsilon^*
| {\dot\epsilon\over\epsilon} + {\mu\over\nu} |^2} }
\exp \left\{-\left[ \frac{ X -  {\nu \over \sqrt{2}}\left[\gamma\epsilon^* 
\left( {\dot\epsilon\over\epsilon} + {\mu\over\nu}\right)^*+
\gamma^* \epsilon
\left( {\dot\epsilon\over\epsilon} + {\mu\over\nu}\right)\right]}
{ \nu \sqrt{ \epsilon^*\epsilon |{\dot\epsilon\over\epsilon} + 
{\mu\over\nu}|^2} }~\right]^2\right\}. \label{wal}
\ee
This expression can be seen to coincide with \eqn{wal2}, hence 
confirming the equivalence of the two approaches.
By evaluating the mean value of $X$ ,
\be
<X> \equiv  <p> \nu+ <q> \mu~,
\ee
we find
\be
<X>= {1\over\sqrt{2}}\left(\gamma \dot\epsilon^* + 
\gamma^*\dot\epsilon\right) \nu +
{1\over\sqrt{2}}\left(\gamma \epsilon^* + 
\gamma^*\epsilon\right) \mu ,
\ee
so that the marginal distribution function for coherent states \eqn{wal} 
takes the simple form
\be
w_\alpha(X,\mu,\nu)= {1\over\sqrt{2\pi\sigma^2_X}} 
\exp\left\{-\frac{(X-<X>)^2}{2\sigma^2_X}\right\},
\ee
where $\sigma^2_X$ is the variance of the variable $X$
\be
\sigma^2_X={1\over 2} \nu^2\epsilon\epsilon^* |{\dot\epsilon\over\epsilon} + 
{\mu\over\nu}|^2~.
\ee

\subsection{The Marginal Distribution Function for n-th Excited States}
From the expression of $w_\alpha$, \eqn{wal2}, we may  evaluate the
MDF, $w_n$, for the n-th excited state. We have 
{
\small
\beqa
w_\alpha(X,\mu,\nu) &=& {1\over \sqrt{\pi|\dot\epsilon\nu + 
\mu\epsilon|^2}} \exp\left\{-|\alpha-\beta|^2-{X^2\over|\dot\epsilon\nu + 
\mu\epsilon|^2}\right\} \cr
&\times& \exp 
\left[-(\alpha-\beta)^2 \frac{(\dot\epsilon^*\nu +\mu\epsilon^*)^2}    
{2|\dot\epsilon \nu + \epsilon \mu|^2} + \sqrt{ 2} (\alpha-\beta) X
\frac{(\dot\epsilon^*\nu +\mu\epsilon^*)}    
{|\dot\epsilon \nu + \epsilon \mu|^2} \right]\cr
&\times& \exp 
\left[-(\alpha-\beta)^{*2} \frac{(\dot\epsilon\nu +\mu\epsilon)^2}    
{2|\dot\epsilon \nu + \epsilon \mu|^2} + \sqrt{ 2} (\alpha-\beta)^* X
\frac{(\dot\epsilon\nu +\mu\epsilon)}    
{|\dot\epsilon \nu + \epsilon \mu|^2} \right],
\eeqa
}
which can be put into the form
{
\small
\beqa
w_\alpha(X,\mu,\nu) & = & 
{ 1\over \sqrt{ \pi}|r |  } 
\exp\left\{ -|\alpha|^2-|\beta|^2-\sqrt{2}\frac{X}{|r|^2}
(r\beta^*+r^*\beta)-\frac{ \beta^2 r^{*2} + \beta^{*2}r^2 + 2X^2}{2 |r|^2} 
\right\} e^{-t^2+ 2tY} e^{ -t^{*2}+ 2t^{*}Y } \cr
& = & {1\over \sqrt{\pi}|r|} 
e^{-|\alpha|^2-Y^2} e^{-t^2+ 2tY} e^{-t^{*2}+ 2t^* Y}
\eeqa
}
with
\bean
r & = & \dot\epsilon\nu + \epsilon\mu \cr
t & = &\alpha\frac{r^*}{\sqrt{2}|r|} \cr
Y & = &\frac{1}{\sqrt{2}}\left[
 \frac{ \beta^*r+  \beta r^* +
\sqrt{2}X }{|r|} \right].
\eean
Remembering the generating functions of Hermite polynomials:
\be
e^{-t^2+ 2tY} = \sum_{n=0}^{\infty} {t^n\over n!} H_n (Y),
\ee
we get (note that $Y$ results to be a real number)
\beqa
w_\alpha(X,\mu,\nu) &=& {1\over \sqrt{\pi}|r|} e^{-|\alpha|^2} e^{-Y^2} 
\sum_{n=0}^{\infty} \sum_{m=0}^{\infty} {1\over n!m!} \cr
& \times & \left(\alpha\frac{r^*}{\sqrt{2}|r|} \right)^n
\left(\alpha^*\frac{r}{\sqrt{2}|r|} \right)^m
H_n\left(Y\right) H_m\left(Y \right).
\eeqa
This in turn must be equal to a series expansion in $w_{nm}$ \cite{new}
\be
w_{\alpha}=  e^{-|\alpha|^2} \sum_{n,m=0}^{\infty} 
\frac{\alpha^n\alpha^{*^m}}{\sqrt{n!m!}} w_{nm} (X,\mu,\nu)~,
\ee
so that we have
\be
w_{nm} (X,\mu,\nu) = {1\over \sqrt{\pi}|r| }
 e^{-Y^2}\frac{1}{\sqrt{n!m!2^{n+m}}}
\left( \frac{r^*}{ |r| } \right)^n
\left( \frac{r}{ |r| } \right)^m
H_n\left( Y \right) H_m\left( Y \right).
\ee
In conclusion the marginal distribution for the n-th excited state results
to be 
\be
w_n(X,\mu,\nu)\equiv w_{nn} (X,\mu,\nu) = 
{ 1 \over \sqrt{\pi}|r| } e^{-Y^2}\frac{1}{n!2^n}
H_n^2\left( Y \right) .
\ee

Let us compare our results with simpler cases.
Assuming $\omega(t)=\omega_0=1$ and $f=0$ we must 
recover the results of ref. \cite{new}. We have
\be
\epsilon(t)=e^{it},~~~\beta(t)=0~,
\ee
consequently
\be
Y=\frac{X}{|r|}=\frac{X}{\sqrt{\mu^2+\nu^2} }
\ee
and we recover the result obtained for the simple harmonic oscillator in 
\cite{new}.

\begin{figure}[]
\epsfxsize=14cm
\epsffile{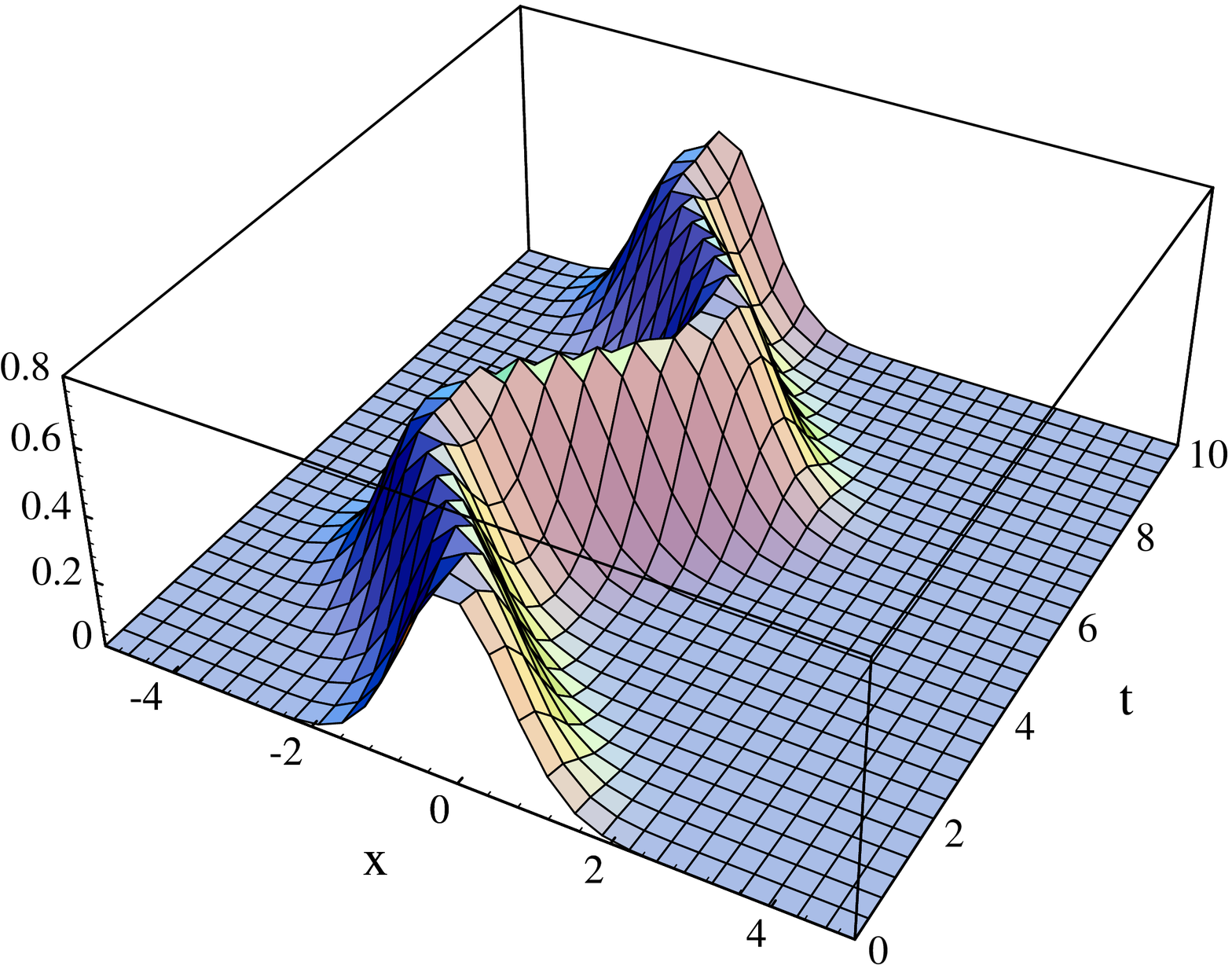}
\caption[]{ $w_0(x,t)$ for $\mu=1$ and $\nu=0$}
\label{f:w0m1n0}
\end{figure}

\begin{figure}[]
\epsfxsize=14cm
\epsffile{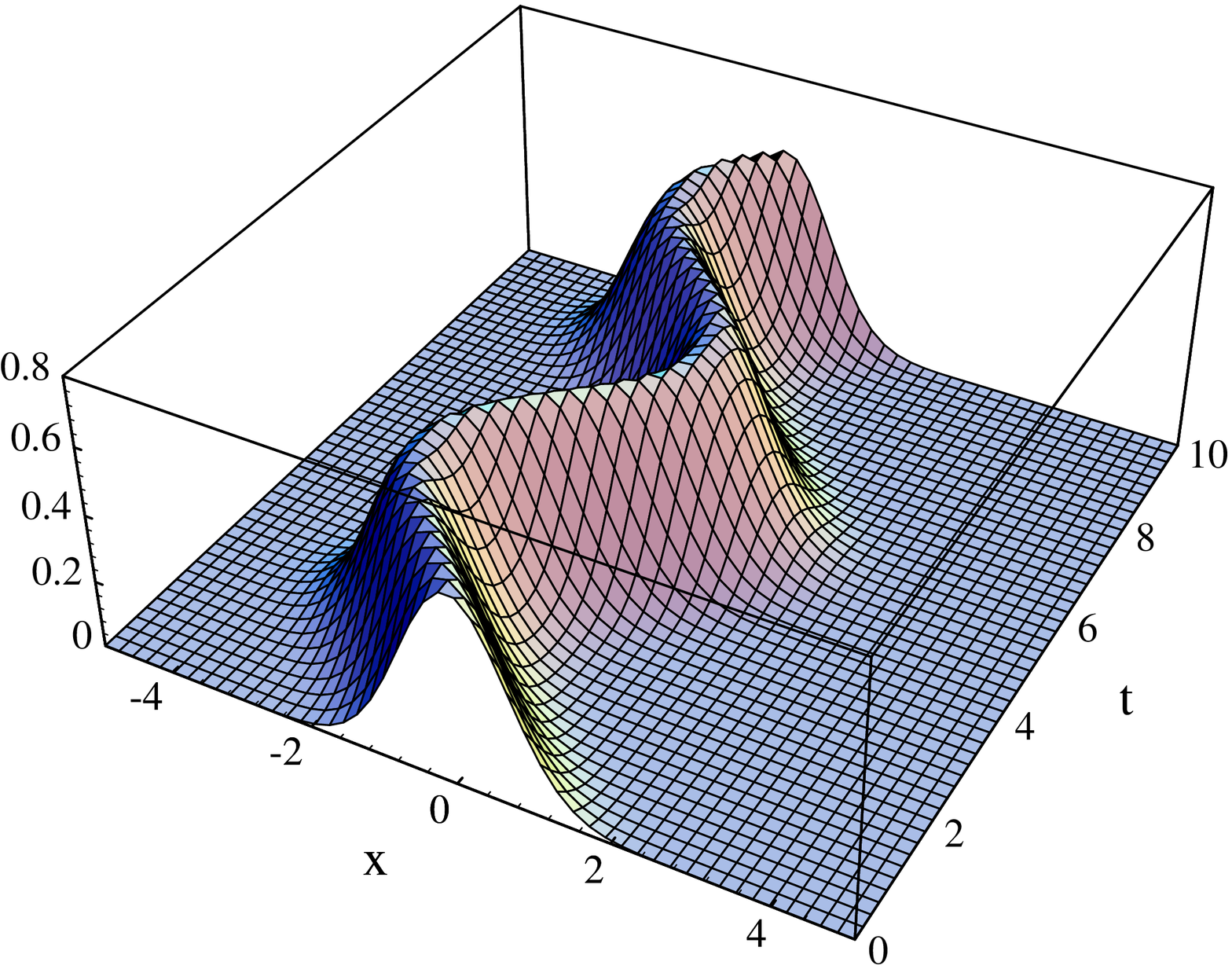}
\caption[]{ $w_0(x,t)$ for $\mu=1/\sqrt{2}$ and $\nu=1/\sqrt{2}$}
\label{f:w0mn05}
\end{figure}

\begin{figure}[]
\epsfxsize=14cm
\epsffile{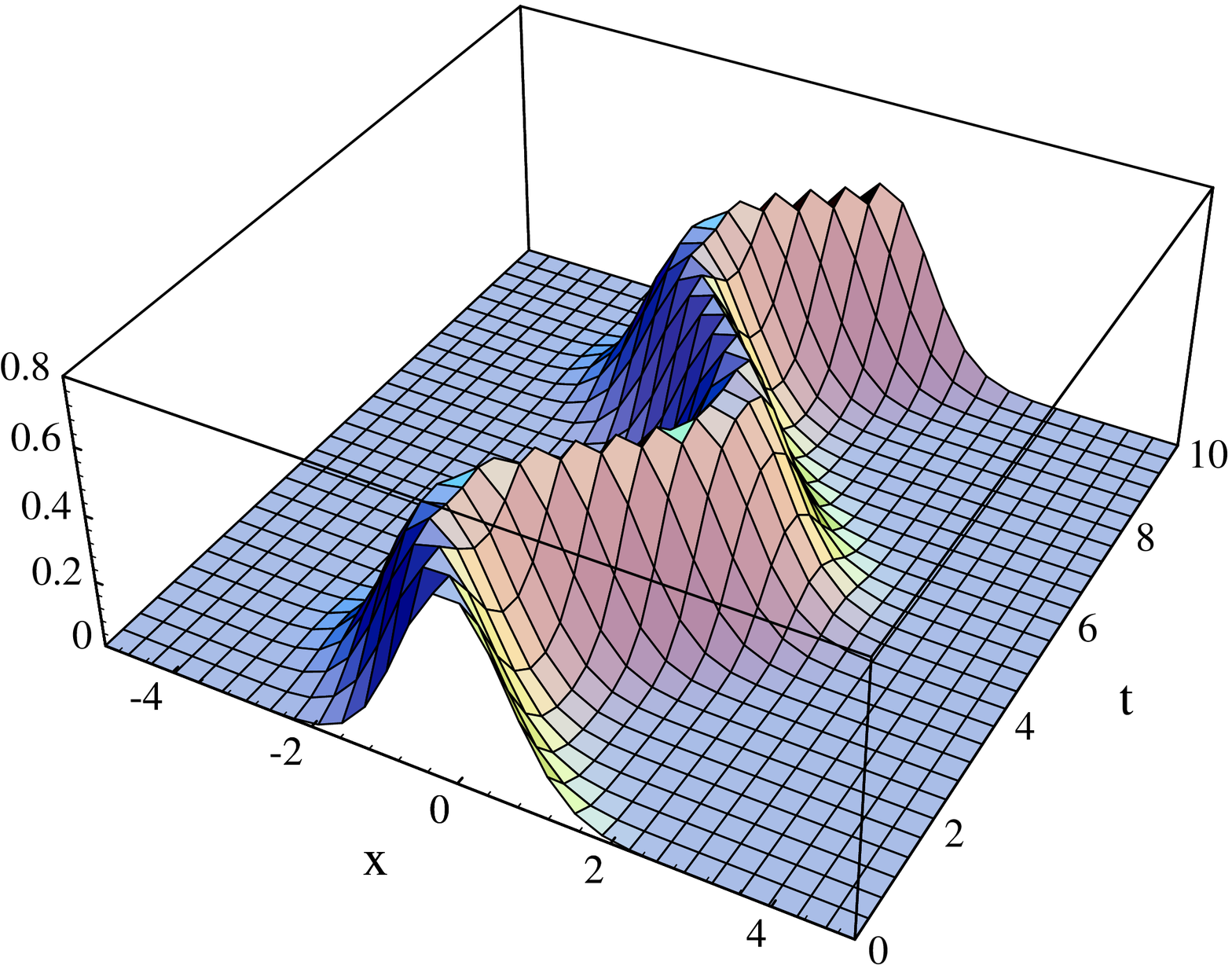}
\caption[]{ $w_0(x,t)$ for $\mu=0$ and $\nu=1$}
\label{f:w0m0n1}
\end{figure}

\begin{figure}[]
\epsfxsize=14cm
\epsffile{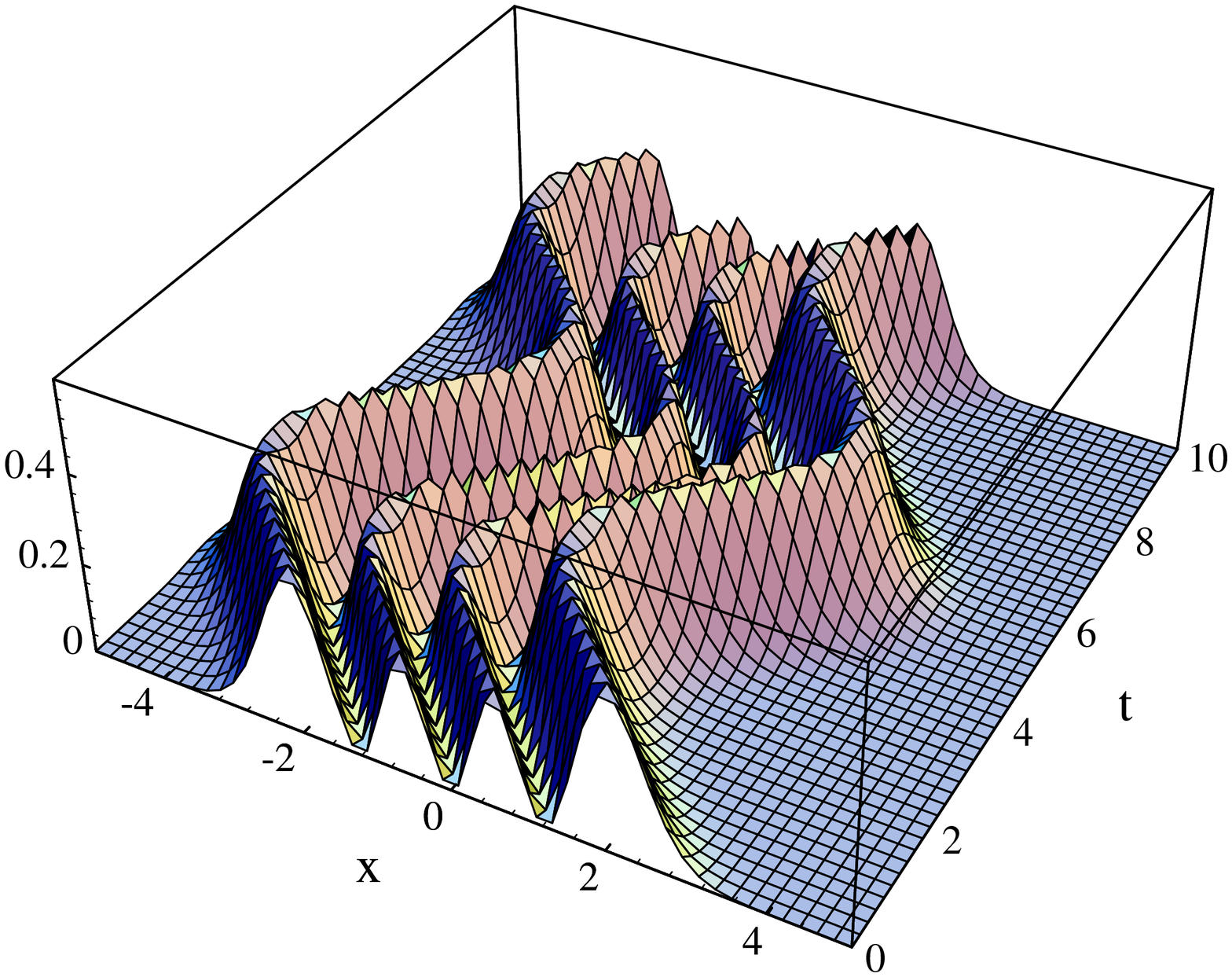}
\caption[]{ $w_2(x,t)$ for $\mu=1/\sqrt{2}$ and $\nu=1/\sqrt{2}$}
\label{f:w3mn05}
\end{figure}

\begin{figure}[]
\epsfxsize=14cm
\epsffile{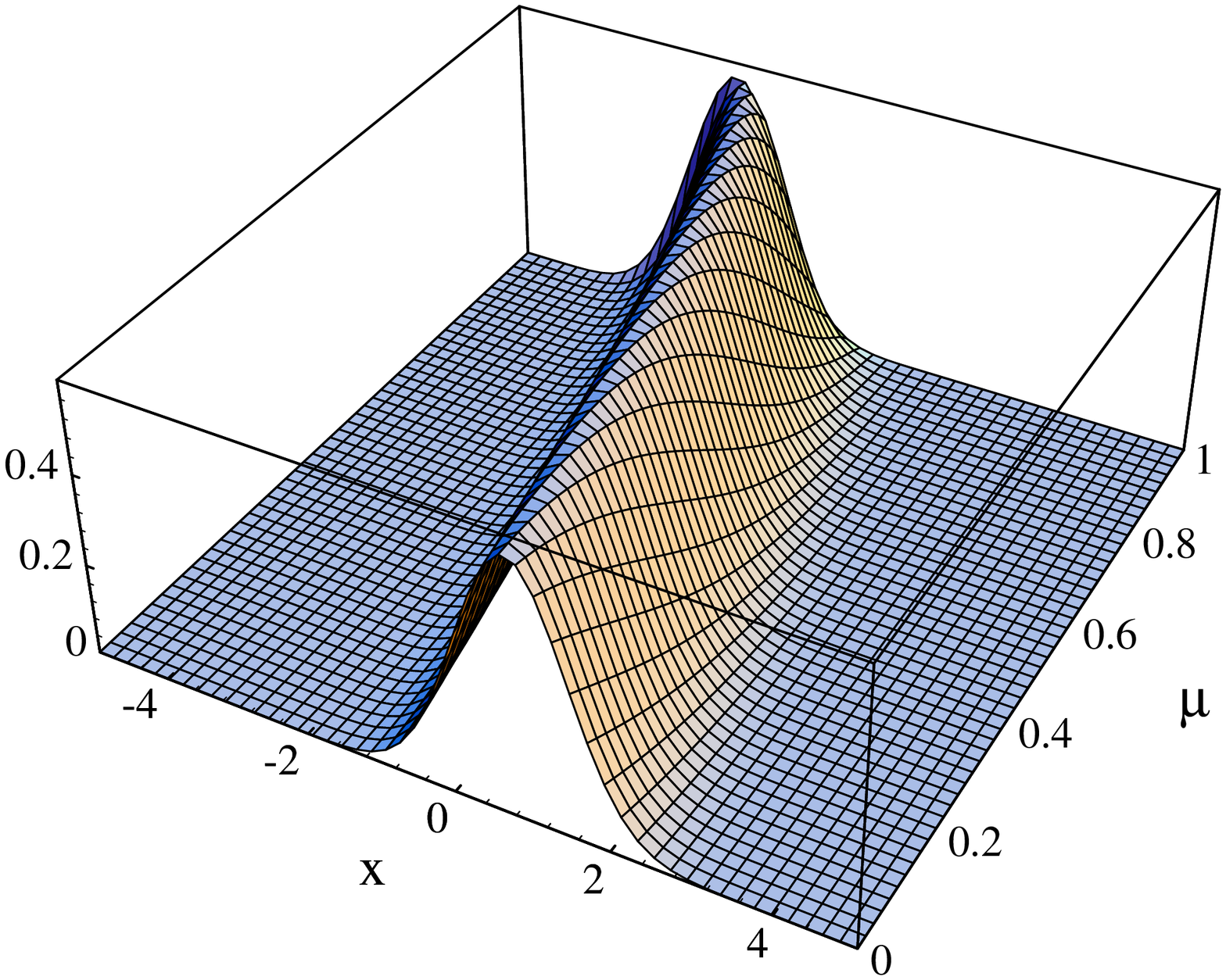}
\caption[]{ $w_0(x,4)$ for $\mu\in(0,1)$ and $\nu=\sqrt{1-\mu^2}$}
\label{f:w0x4}
\end{figure}

\begin{figure}[]
\epsfxsize=14cm
\epsffile{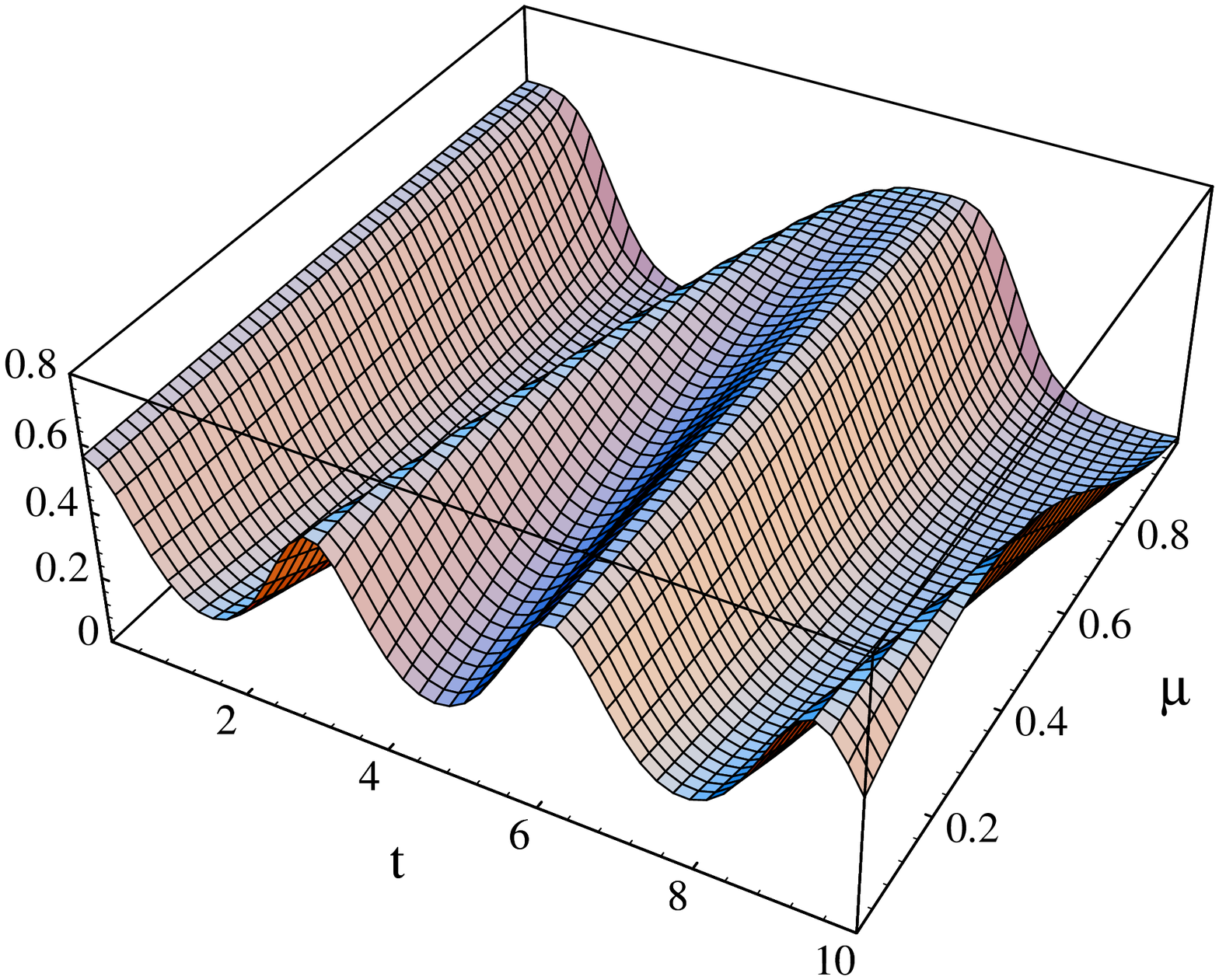}
\caption[]{ $w_0(0,t)$ for $\mu\in(0,1)$ and $\nu=\sqrt{1-\mu^2}$}
\label{f:w00t}
\end{figure}

To give an idea of what happens if $\omega(t)\neq$ const. we show
below some plots of $w_n(x,t)$ for different values of the parameters.
In this respect it is necessary to solve Eq.\eqn{epsi} for
$\epsilon(t)$. In the parametric resonance case (in dimensionless
units) 
\be
\omega^2(t)=\frac{1+k\cos{2t}}{1+k},~~~k\ll 1
\ee 
a good approximation for $\epsilon(t)$ is given by
\cite{domani}  (see also \cite{domaro}):
\be
\epsilon(t)=\cosh{\frac{k t}{4}}e^{it}-i\sinh{\frac{kt}{4}}e^{-it}.
\ee
Thus, in this case, $w_n(x,t)$ is completely determined.
Fig.\ref{f:w0m1n0} shows the plot of $w_0(x,t)$ when $\mu=1,~\nu=0$ and
$k=0.01$. Since in this case $H_0(Y)=1$ we obtain the typical Gaussian
centered around the origin of the axes, modulated in time by the factor
${1}/{|r|}=(1-\sinh{\frac{k t}{2}}\sin{t} )^{-1/2} $. When $n$
increases we recognize the appearance of the typical structure of
minima and maxima due to the zeroes of the Hermite polynomials
Fig.\ref{f:w3mn05}. Quite interesting is the behaviour of $w_n(x,t)$
with respect to $\mu$ and $\nu$, which can be seen in
Fig.\ref{f:w0mn05} and Fig.\ref{f:w0m0n1} (see also Fig.\ref{f:w0x4}
and Fig.\ref{f:w00t}). To study the dependence of $w$ on $\mu$ and
$\nu$ we must remember that the information contained in the MDF is
over-complete in the sense that we can choose different ``tomography
schemes'' \cite{new}. This is the counterpart of the different
representations (coordinate, momentum etc.) existing in the usual
formulation of quantum mechanics. In the following we use the optical
tomography scheme \cite{vogrisk,raymer}: $\mu^2+\nu^2=1$. In Fig.
\ref{f:w0x4} we show $w_0$ as a function of $x$ and $\mu$, we note that
the maximum of probability shifts from $x>0$ to $x<0$ when $\mu$ goes
from 0 to 1 corresponding to the two extreme cases $X=p$ and $X=q$
respectively. Fig. \ref{f:w00t} shows the change of $w$ at a fixed
point $x$ as a function of $t$ and $\mu$. We observe a typical
oscillation behaviour at fixed $\mu$ for varying time and at a fixed
instant we note the change of $w$ with respect to $\mu$. 

\section{Concluding remarks}
In this paper we studied the driven harmonic oscillator in the
framework of the probability representation of quantum mechanics. By
means of the time-dependent invariants we determine the classical
propagator, $\Pi$, of the evolution equation of the marginal distribution
function relative to the potential considered. In this way we were able
to reconstruct the quantum propagator for the density matrix and, up to
a phase factor, the quantum propagator for the wave function. We
recover well known limit cases \cite{new} and \cite{feynman}. 

We compute the marginal distribution function, $w_\alpha$, for coherent
states. We obtain it first in the framework of the probability
representation, then using the usual techniques of quantum mechanics.
The time dependence of $w_\alpha$ is achieved both by means of the
classical propagator and, by directly solving a time dependent equation
which encodes the notion of coherence. 

Starting from $w_\alpha$ we compute the marginal distribution
eigenfunctions in the energy eigenstate basis and study its behaviour
in a particular case (the parametric resonance). 

Our results drive us to the conclusions that the quantum description of
the forced parametric oscillator can be given in a selfconsistent
approach, which is alternative to the Schr\"odinger picture while
closer to the classical description. This is a further evidence for the
possibility of formulating quantum mechanics by means of the MDF
associated to a random variable $X$, avoiding complex wave functions
and the density matrix formalism. 



\newpage

\newpage

\newpage

\newpage

\newpage

\newpage

\end{document}